\documentclass[ reprint, amsmath, amssymb, aps, pra, superscriptaddress ]{revtex4-2}

\usepackage{graphicx}
\usepackage{dcolumn}
\usepackage{bm}
\usepackage{epstopdf}
\usepackage{braket}
\usepackage{tcolorbox}
\usepackage[breaklinks=true,colorlinks]{hyperref}
\begin{document}

\title{{Supersensitive phase estimation by thermal light in a Kerr-nonlinear interferometric setup}}


\author{Nilakantha Meher}
\email{nilakantha.meher6@gmail.com}
\affiliation{Department of Chemical and Biological Physics, Weizmann Institute of Science, Rehovot 7610001, Israel}
\affiliation{Department of Physics, SRM University-AP, Amaravati 522240, Andhra Pradesh, India}
\author{Eilon Poem}
\email{eilon.poem@weizmann.ac.il}
\affiliation{Department of Physics of Complex Systems, Weizmann Institute of Science, Rehovot 7610001, Israel}
\author{Tom\'{a}\v{s} Opatrn\'y}
\affiliation{Department of Optics, Faculty of Science, Palack\'y University, 17. listopadu 50, 77146 Olomouc, Czech Republic}
\author{Ofer Firstenberg}
\affiliation{Department of Physics of Complex Systems, Weizmann Institute of Science, Rehovot 7610001, Israel}
\author{Gershon Kurizki}
\affiliation{Department of Chemical and Biological Physics, Weizmann Institute of Science, Rehovot 7610001, Israel}



\begin{abstract}
{Estimation of the phase delay between interferometer arms is the core of transmission phase microscopy. Such phase estimation may exhibit an error below the standard quantum (shot-noise) limit, if the input is an entangled two-mode state, e.g., a N00N state. We show, by contrast, that such supersensitive phase estimation (SSPE) is achievable by \textit{incoherent}, e.g., \textit{thermal}, light that is injected into a Mach-Zehnder interferometer via a Kerr-nonlinear two-mode coupler. Phase error is shown to be reduced below $1/\bar{n}$, $\bar{n}$ being the mean photon number, by thermal input in such interferometric setups, even for small nonlinear phase-shifts per photon pair or for significant photon loss. Remarkably, the phase accuracy achievable in such setups by thermal input surpasses that of coherent light with the same $\bar{n}$.} Available mode couplers with giant Kerr nonlinearity that stems either from dipole-dipole interactions of Rydberg polaritons in a cold atomic gas, or from cavity-enhanced dispersive atom-field interactions, may exploit such effects to substantially advance interferometric phase microscopy using incoherent, faint light sources. 
\end{abstract}

\date{\today}


\maketitle

\section{Introduction}
Quantum sensing generally denotes sensing protocols that exploit nonclassical resources or operate in the domain of quantum signals and noise \cite{Degen2017RevModPhys, Pezze2018RevModPhys, Kurizki2015PNAS, KurizkiBook, Dowling2003quantum, Zwick2016PRAppl,fatemi2017modal, Anderson2023PhysToday}. One of the prominent directions in this emerging field is that of transmission microscopy with phase resolution set by super-sensitive phase measurements, namely, measurements with uncertainty less than the standard quantum limit (SQL) \cite{Ou1997PRA,Caves1981Quantum-mechanicalInterferometer,Lang2013PRL,Buzek1995QuantumMeasurements,Hofmann2007High-photon-numberLight}. In conventional two-mode interferometers, the SQL is set by the vacuum noise in the empty input mode \cite{Takeoka2017PRA}, corresponding to the shot-noise error of phase-delay estimation $\Delta\phi$ that scales as $\Delta\phi \sim \bar{n}^{-1/2}$, $\bar{n}$ being the mean input photon number in the populated mode. 

{Phase estimation below SQL, alias supersensitive phase estimation (SSPE), is enabled by nonclassical two-mode input states, \textit{e.g.} squeezed vacuum states \cite{Lang2013PRL,Buzek1995QuantumMeasurements,Kalinin2023Nanophotonics}, even- and odd-coherent states \cite{Luis2001PRA, meher2020QINP}, N00N states \cite{Hofmann2007High-photon-numberLight} and entangled coherent (\textquotedblleft cat\textquotedblright) states (ECS) that consist of coherently superposed N00N states  \cite{Luis2001PRA}. Both N00N states and ECS yield the  phase-error bound $\Delta\phi \sim 1/\bar{n}$, which is commonly but inaccurately identified with the Heisenberg limit (HL) \cite{Ou1997PRA,Caves1981Quantum-mechanicalInterferometer,Buzek1995QuantumMeasurements,Luis2001PRA,Lang2013PRL,Hofmann2007High-photon-numberLight,Joo2011QuantumStates, Milburn1986QuantumOscillator,Tombesi1987GenerationTechnique,Sanders1992EntangledStates,meher2020QINP}. 
However, the states with small $\bar{n}$ and large variance $\Delta n>\bar{n}$ may exhibit phase error below $1/\bar{n}$, reflecting the fact that the HL as the ultimate limit for phase estimation does not apply for such states \cite{Ou1997PRA,Monras2006PRA,Hofman2009PRA,Pinel2013PRA,Boixo2007PRL}.} 

To benefit from the input state nonclassicality, SSPE must be inferred from high-order correlation measurements of the two-mode output \cite{Campos2003OpticalMeasurements}. Progress over the years \cite{Gerry2002NonlinearInterferometry, Paternostro2003GenerationRegime, Ourjoumtsev2007GenerationStates,Takahashi2008GenerationSubtraction, Obrien2009NatPhot, Afek2010High-NOONLight,raizen1987squeezed,
Gerrits2010GenerationVacuum, Krischek2011UsefulEstimation,Ou2012EnhancementInterferometer, Rosen2012SubRayleighLossResistant, Israel2014SupersensitiveLight, Hudelist2014QuantumInterferometers, Feizpour2015NonlinearPhaseByPostSelection,  Wang2016ABoxes,Wang2016ABoxes, Israel2019EntangledCoherentStates, Genovese2021AVSQS, Firstenberg2016NonlinearInteractions, Firstenberg2013AttractiveMedium,
Drori2023Sc, birrittella2021parity} has revealed the hurdles involved in the generation of such states: especially challenging are the giant nonlinear phase shifts of $ \pi$ per photon pair required for their deterministic preparation,  as well as their fragility under decoherence.  

{Here we break away from the underlying paradigm that SSPE must employ interferometers with maximally entangled states. We show that \textit{nonclassical or coherent sources are not required} for deterministic SSPE. Instead, it is achievable by feeding a Mach-Zehnder interferometer (MZI) that constitutes the core of a transmission phase microscope with light derived from \textit{incoherent, e.g., thermal, input} shined upon another Mach-Zehnder interferometer with Kerr-nonlinear two-mode coupler.} Not less remarkably, the nonlinear phase shift per photon pair required to beat the SQL limit of phase microscopy can be much less than the $\pi$-shift which transforms a Fock state into a N00N state or a coherent state into an ECS \cite{Sanders1992EntangledStates}. Such Kerr-nonlinear SSPE is also impervious to decoherence and can withstand substantial photon loss inside the interferometer. The discussed effects may be experimentally realized by resorting to the strong optical Kerr effects predicted \cite{Friedler2005PRA, Shahmoon2011PRA} and demonstrated in setups based on dipole-dipole interacting Rydberg polaritons in a cold gas \cite{Firstenberg2016NonlinearInteractions,Firstenberg2013AttractiveMedium,
Drori2023Sc, Adams2019JPHYB, Obrien2009NatPhot}, or atom-field interactions in high-$Q$ cavities \cite{stolz2022quantum, Bechler2018NatPhys}.  
\begin{figure*}
\begin{center}
\includegraphics[height=9.5cm,width=16cm]{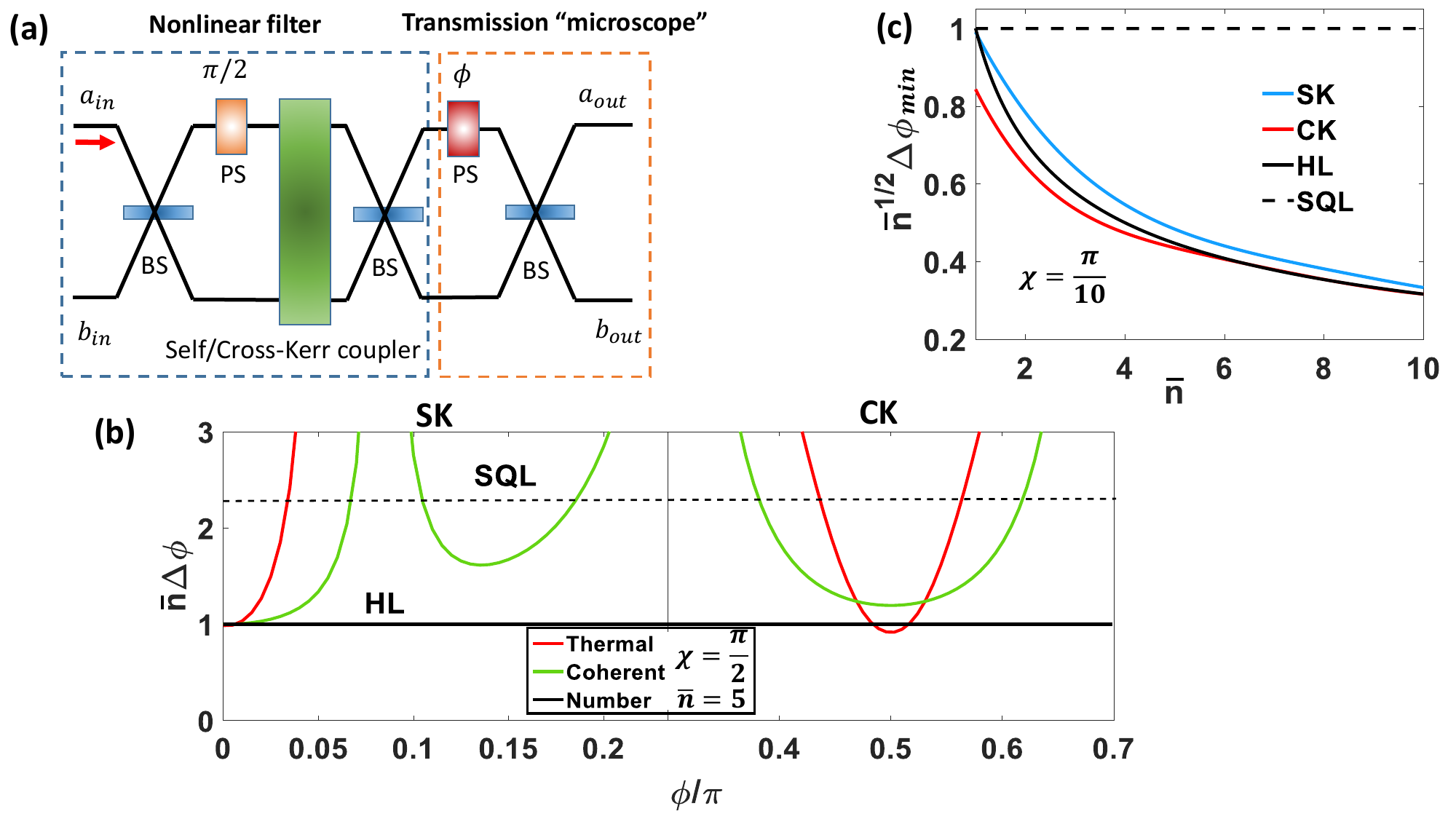}
\end{center}
\caption{{(a) Schematic drawing of a Mach-Zehnder interferometer (MZI) preceded by either a self-Kerr (SK) or a cross-Kerr (CK) nonlinear filter (two-mode coupler). The nonlinear filter of thermal input in modes $a_{in},b_{in}$, prepares states allowing to estimate with high precision the unknown phase shift (PS) $\phi$ in a linear MZI. The setup can act as transmission \textquotedblleft microscope" (phase resolver).} The phase is inferred from photodetection of the output channels $a_{out},b_{out}$ or  parity measurement of one of them. (b) The phase error (sensitivity) from parity measurements as a function of $\phi$ for various input states in the SK and CK MZIs (with $\chi=\pi/2,\bar{n}=5$). (c) The minimal phase error (maximum sensitivity), given by the inverse square root of the quantum Fisher information (QFI), as a function of $\bar{n}$ for thermal state input to the MZI with weak SK or CK nonlinearity ($\chi=\pi/10$). }\label{SetupParityDetect}
\end{figure*} 

\section{Scenario and analysis principles}
 Consider a two-mode interferometer that incorporates a phase-shifter with an unknown phase $\phi$ that we wish to estimate. The two-mode merger/coupler is endowed with self-Kerr (SK) or cross-Kerr (CK) nonlinearity [Fig. \ref{SetupParityDetect}(a)]: A nonlinear phase shift $2\chi$ is caused by SK for two photons in the same mode, the CK counterpart being the phase shift $\chi$ for two photons in different modes. The overall unitary transformation for such a Mach-Zehnder interferometer (MZI) with SK or CK nonlinearity \cite{chekhova2016nonlinear, kitagawa1986number} is, respectively,
\begin{subequations}
\begin{align}
U_{SK}&=U_{BS} U_{PS}(\phi) U_{BS} e^{i\chi a^{\dagger 2} a^2} U_{PS}(\pi/2) U_{BS},\\
U_{CK}&=U_{BS} U_{PS}(\phi) U_{BS} e^{i\chi a^\dagger a b^\dagger b} U_{PS}(\pi/2) U_{BS}.
\end{align}
\end{subequations}
Here, $U_{PS}(\phi)=e^{i\phi a^\dagger a}$ is the operator that shifts the phase by $\phi$ in mode $a$, and $U_{BS}$ is the operator of a  50:50 beam splitter \cite{Gerry}. Both Hamiltonians $H_{SK}=\chi a^{\dagger 2} a^2$ and $H_{CK}=\chi a^\dagger a b^\dagger b$ commute with the photon-number operators $a^\dagger a$ and $b^\dagger b$ and thus, do not change the mean output photon numbers \cite{Imoto1985PRA}. We consider the simplest (both experimentally and calculationally) case of one empty input mode and the other mode to be in either number, coherent or thermal states \cite{Gerry,Scullybook97}. Analytical results will be given for the strongest nonlinear phase-shift per photon pair, $\chi=\pi/2$. 

An insightful (but not optimal) phase estimation method is based on measuring the mean value of the parity operator of one of the output modes: $\Pi_{out}=e^{i\pi b^\dagger_{out}b_{out}}$ \cite{Gerry2002NonlinearInterferometry,birrittella2021parity}. Its direct measurement by counting the corresponding photon number is challenging. Instead, Kerr nonlinearity enables a simpler parity measurement of an MZI output by sending it through a second cross-Kerr interferometer along with a vacuum input and detecting which of the two output detectors has clicked [see Fig. \ref{ParityFilter}] \cite{Gerry2002NonlinearInterferometry}. The phase error (sensitivity) obtained from the parity-operator mean value $\langle \Pi_{out} \rangle$ is compactly expressed as \cite{Gerry2002NonlinearInterferometry,birrittella2021parity}
\begin{align}\label{Sensitivity}
\Delta\phi=\frac{\sqrt{1-\langle \Pi_{out} \rangle^2}}{\left|\frac{\partial \langle \Pi_{out} \rangle}{\partial \phi}\right|}.
\end{align}

{In an MZI with self-Kerr (SK) and $\chi=\pi/2$, for number state input $\ket{n,0}$, the state right after the phase shifter [Fig. \ref{SetupParityDetect}(a)], is
\begin{align}\label{NumberInputPsiphi}
\ket{\psi(\phi)}_{number}=\frac{i^n}{\sqrt{2}}e^{i\pi/4}(i^{n-1}e^{in\phi}\ket{n,0}+\ket{0,n}),
\end{align} 
which yields at the output of the microscope $\langle \Pi_{out} \rangle=\sin n\phi$ [see Appendix \ref{A}] and $\Delta\phi=1/n$ (Heisenberg scaling)}. One owes this supersensitive phase resolution to the fact that a number state is transformed after the second beam splitter/merger in Fig. \ref{SetupParityDetect}(a) into a state  akin to a N00N state.  Accordingly, an input thermal state is  transformed by the nonlinear filter into a mixture of N00N states spanned by all $n$
which yields 
\begin{align}\label{PavgThermal}
\langle \Pi_{out}\rangle= \frac{1}{1+\bar{n}}+\sum_{n=0}^\infty \frac{\bar{n}^n}{(1+\bar{n})^{(n+1)}}\sin n\phi.
\end{align}
The resulting minimum phase error [see Fig. \ref{SetupParityDetect}(b) and Appendix \ref{A}] 
\begin{align}\label{DphiThermal}
\Delta\phi\approx\frac{\sqrt{1-\frac{1}{(\bar{n}+1)^2}}}{\bar{n}},
\end{align}
\textit{falls below the Heisenberg limit (HL)} $1/\bar{n}$ \cite{Ou1997PRA,Monras2006PRA,Hofman2009PRA,Pinel2013PRA,Boixo2007PRL}. Importantly, similar arguments hold for a general input state $\rho_{in}=\sum_{n,m}\rho_{nm} \ket{n}\bra{m} \otimes \ket{0} \bra{0}$ [see Appendix \ref{A}].  We note that there is only moderate difference in phase sensitivity between cross-Kerr (CK) and SK, all parameters being equal [Fig. \ref{SetupParityDetect}(b)]. 

Independently of the measurement setting, the minimal phase error attainable by a given input state is set by the quantum Cram{\'e}r-Rao bound \cite{cramer1999mathematical}, which is related to the quantum Fisher information (QFI) $F_Q$ through \cite{birrittella2021parity}
\begin{align}\label{CramerRao}
\Delta\phi_{min} \geq \frac{1}{\sqrt{F_Q}}.
\end{align} 

{To evaluate $F_Q$, we calculate the state right after the phase-shift operation [Fig. \ref{SetupParityDetect}(a)]  through
\begin{subequations}\label{StateforFQ}
\begin{align}
\tilde{\rho}(\phi)=U\rho_{in}U^\dagger,
\end{align}
where 
\begin{align}
U&=U_{PS}(\phi) U_{BS} e^{i\chi a^{\dagger 2} a^2} U_{PS}(\pi/2) U_{BS},~~~ \text{for SK}\\ 
U&=U_{PS}(\phi) U_{BS} e^{i\chi a^\dagger a b^\dagger b} U_{PS}(\pi/2) U_{BS},~~~\text{for CK}.
\end{align}
 \end{subequations}}
{Using Eqs. \eqref{NumberInputPsiphi} and \eqref{StateforFQ}, the transformed states after the phase-shifter in SK interferometer with  $\chi=\pi/2$ for coherent and thermal states input correspondingly are calculated to be [see Appendix \ref{A}]
\begin{subequations}\label{TransformedStates}
\begin{align}
&\ket{\psi({\phi})}_{coherent}= e^{\frac{-|\alpha|^2}{2}}\sum_{n=0}^\infty\frac{(i\alpha)^n}{\sqrt{2}\sqrt{n!}}(i^{n-1}e^{in\phi}\ket{n,0}+\ket{0,n}),\\
&\tilde{\rho}_{thermal}(\phi) =\frac{1}{2}\sum_{n=0}^\infty \frac{\bar{n}^n}{(1+\bar{n})^{(n+1)}}\left[\ket{n,0} \bra{n,0}+\ket{0,n} \bra{0,n} \right.\nonumber\\
&~~~~~~~~+ \left. i^{n-1}e^{in\phi}\ket{n,0}\bra{0,n}+  (-i)^{n-1}e^{-in\phi}\ket{0,n}\bra{n,0} \right].
\end{align} 
\end{subequations}
Similarly, one can obtain the transformed states for CK interferometer.}

{For $\tilde{\rho}(\phi)$, the QFI is given by the formula \cite{birrittella2021parity,Toth2014JPA}
\begin{align}\label{FisherInf}
F_Q(\tilde{\rho})=2\sum_{\substack{k,l \\ \lambda_k + \lambda_l > 0}}\frac{(\lambda_k-\lambda_l)^2}{(\lambda_k+\lambda_l)}|\langle k|a^\dagger a|l\rangle|^2,
\end{align}
where $\{\lambda_k, \ket{k}\}$ are the eigenvalues and their corresponding eigenvectors of $\tilde{\rho}(\phi)$.} {For a pure state input \cite{Lang2013PRL,Lang2014PRA}, Eq. \eqref{FisherInf} reduces to 
\begin{align}\label{FisherInfPure}
F_{Q}=4(\langle  \psi(\phi) |(a^\dagger a)^2 | \psi(\phi)\rangle-|\langle  \psi(\phi)| a^\dagger a|\psi(\phi) \rangle|^2)=4\langle \Delta a^\dagger a\rangle^2,
\end{align}
where $| \psi(\phi)\rangle=U\ket{\psi_{in}}$.}

Using Eqs. \eqref{NumberInputPsiphi}, \eqref{FisherInf}, \eqref{FisherInfPure} and \eqref{TransformedStates}, we obtained the following $F_Q$ for thermal, coherent and number states input, which satisfy a remarkable analytical \textit{inequality} [see Appendix \ref{C}]
\begin{widetext}
\begin{subequations}\label{FQforAll}
\begin{align}
(F_Q)_{thermal}=2 \bar{n}^2+\bar{n} > (F_Q)_{coherent}=\bar{n}^2+2\bar{n} > (F_Q)_{number}=\bar{n}^2,~~~ \text{for SK}\\
(F_Q)_{thermal}= \bar{n}^2+\bar{n}  > (F_Q)_{number}=\bar{n}^2 > (F_Q)_{coherent}=\tfrac{1}{2}\bar{n}^2+2\bar{n}, ~~~ \text{for CK}
\end{align}
\end{subequations}
\end{widetext}
with  $\bar{n}> 1$ in Eq. (\ref{FQforAll}a) or $\bar{n}> 4$ in Eq. (\ref{FQforAll}b). This inequality shows that the minimal phase error $\Delta\phi_{min}=1/\sqrt{F_Q}$ is below the HL $1/\bar{n}$ for thermal input,  and that \textit{thermal input allows higher phase sensitivity than coherent- and number-state inputs}. The phase sensitivity, which is determined by the QFI, depends on the second moment of the photon number of the probe state [see Appendix \ref{C}]. For a given mean number of photons $\bar{n}$, the state with the largest  $\langle a^{\dagger 2}a^2\rangle$ exhibits maximum QFI. Since the probe state (after second beam splitter) generated from thermal input has $\langle a^{\dagger 2}a^2\rangle$ larger than that of the coherent-state input, thermal input yields, surprisingly, larger QFI.

SK outperforms CK for $\chi=\pi/2$ in terms of QFI, since all number states input are transformed to N00N states for SK, whereas only odd number states do so for CK [see Appendix \ref{A}]. Surprisingly, numerical calculations of the minimal phase error $(1/\sqrt{F_Q})$, for weak $\chi\simeq \pi/10$ and small average photon number $\bar{n}\simeq 1$, show that CK can beat the HL and outperforms SK, and that both allow beating the SQL [see Fig. \ref{SetupParityDetect}(c)]. This supersensitivity achieved for $\chi\leq 1$ although N00N states are then not formed.

\begin{figure}
\begin{center}
\includegraphics[height=6.5cm,width=9cm]{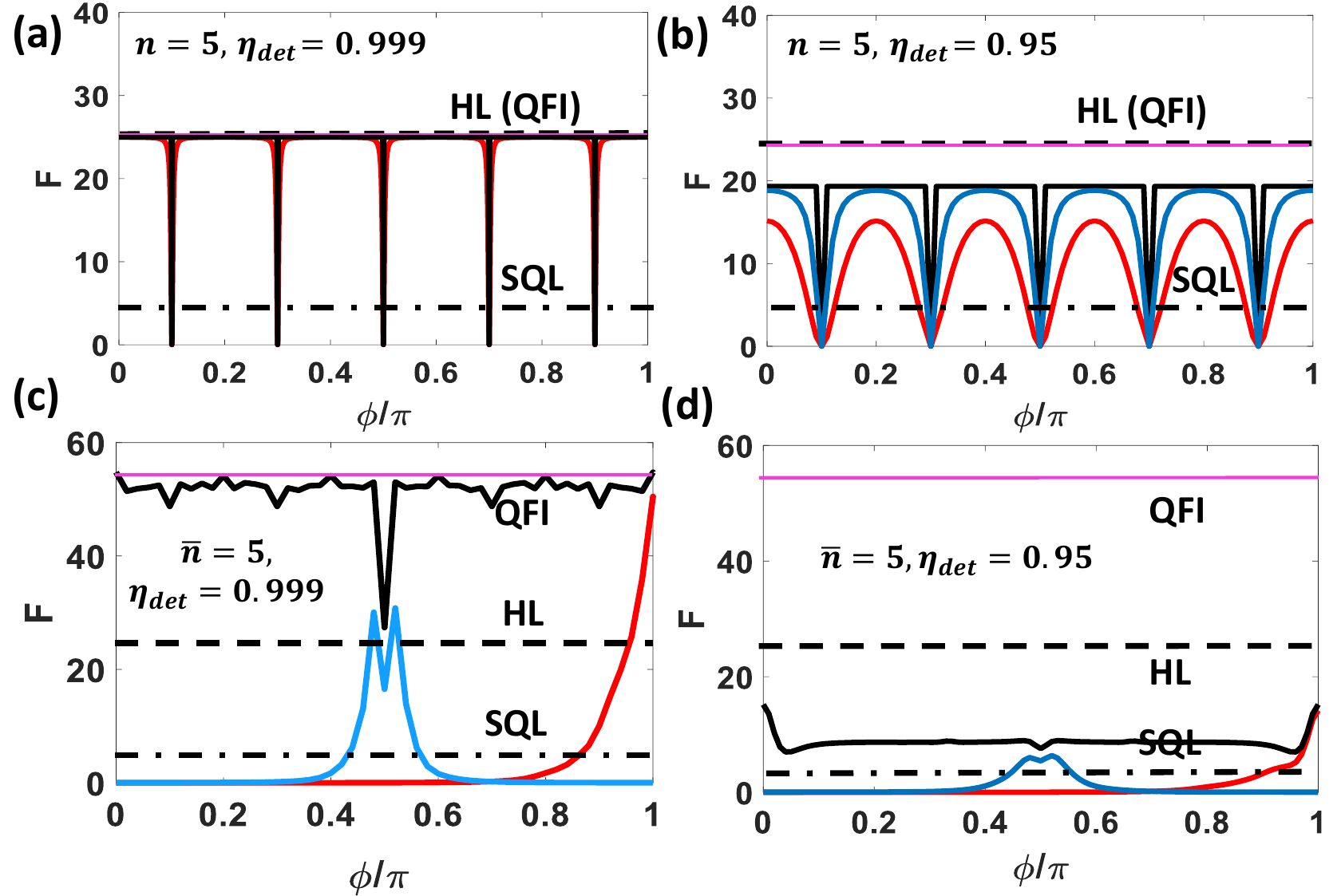}
\end{center}
\caption{ {Fisher information for SK-filtered MZI ($\chi=\pi/2$) obtained by photon-number resolving detectors: number-state ($n=5$) input with detector efficiencies (a) $\eta_\text{det}= 0.999$, (b) $\eta_\text{det}=0.95$, and thermal state ($\bar{n}=5)$ input with detector efficiencies (c) $\eta_\text{det}= 0.999$, (d) $\eta_\text{det}=0.95$. The SQL (dot-dashed line) here corresponds to $F_{SQL}=\bar{n}=5$ and the HL (dashed line) to $F_{HL}=\bar{n}^2=25$. The QFI (magenta line) for thermal input is 55 and for number input is 25. Red thick line: Single-detector photodetection, Blue thick line: intensity-difference detection, Black thick line: photodetection of both detectors. For $\eta_\text{det}\geq 0.95$, thermal input yields $F\geq 10$ and thus beats the SQL.  }  }\label{FisherNumThermInEff}
\end{figure} 


\noindent
\textbf{Photodetection efficiency:}
Numerical investigations of the Fisher information (Fig. \ref{FisherNumThermInEff}) for phase estimation, obtained either from the count difference of the two photon-number resolving detectors of $a_{out}$ and $b_{out}$ in Fig. \ref{SetupParityDetect}(a) or from their complete data, yield the following results for nonlinearity $(\chi=\pi/2)$ and imperfect detectors with efficiency $\eta_\text{det}$: (a) The Fisher information $(F)$ based on complete data from both detectors is always higher than or equal to the Fisher
information based on partial data. (b) A decrease in the detection efficiency leads to a fast drop of $F$ and its more pronounced dependence on $\phi$.  (c) With perfect detectors $\eta_\text{det}=1$, photodetection on both output channels saturates the QFI limit for thermal- and number-state inputs [Fig. \ref{FisherNumThermInEff}], but not for coherent state input [see Appendix \ref{D}]. With near-perfect detectors $\eta_\text{det}\geq 0.97$, mean photon numbers being equal, the highest $F$ is reached by thermal states, which then beat the $1/\bar{n}$ limit and  surpass coherent and Fock states consistently with the inequalities in Eq. \eqref{FQforAll}. For  $\eta_\text{det}\leq 0.97$ the order is reversed: the highest $F$ is reached by Fock states and the lowest by thermal. Yet, thermal input still beats the SQL for $\eta_{\text{det}} \geq 0.9$ [Fig. \ref{ThermalDetEff09}]. We note that the FI based on parity measurement reaches the QFI for CK MZI and not for SK MZI [see Appendix \ref{D}].

\begin{figure}
\begin{center}
\includegraphics[height=7cm,width=9cm]{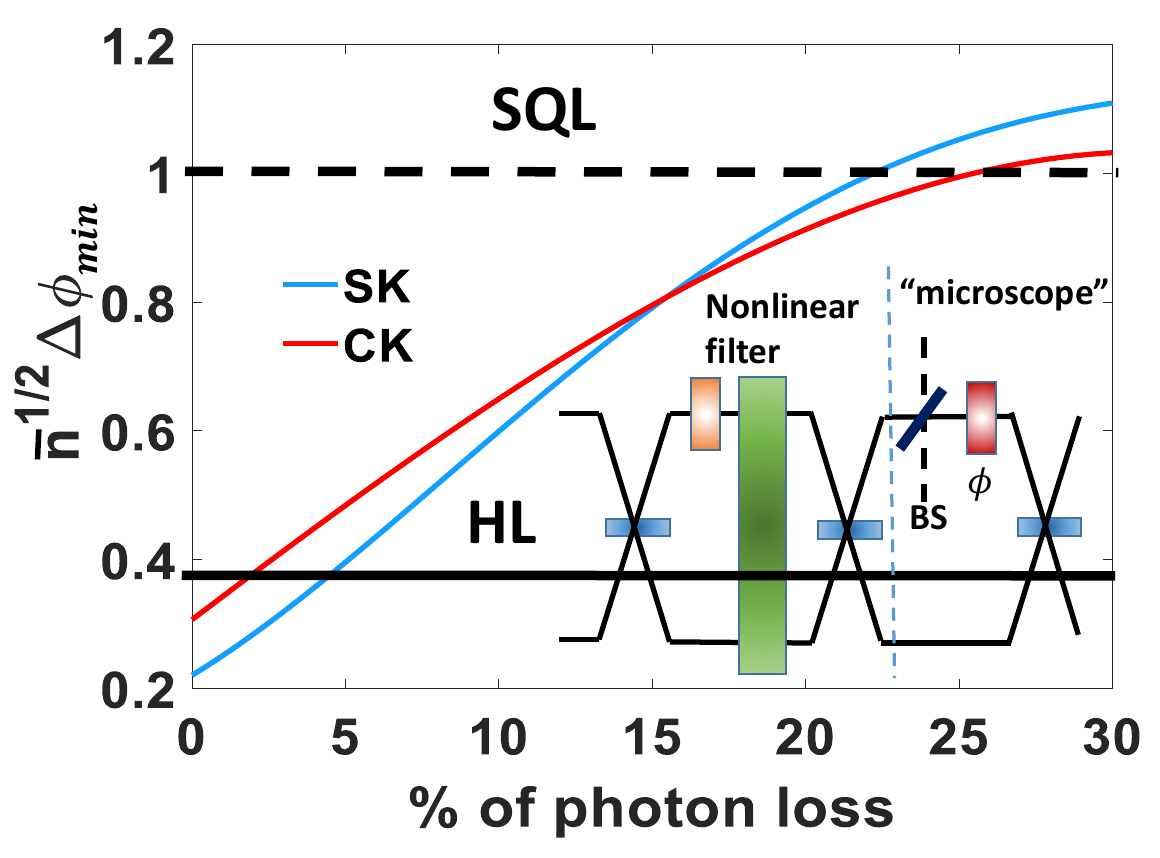}
\end{center}
\caption{Inset: Lossy Kerr-nonlinear MZI mimicked by a fictitious beam splitter in one arm. Plot: Minimal phase error, calculated for thermal state from Eq. (\ref{CramerRao}), as a function of the percentage of photon loss in one arm of either CK or SK MZI. Here, $\bar{n}=10$ and $\chi=\pi/2$.}
\label{EntanglementPhotonLoss}
\end{figure}


\noindent
\textbf{Photon loss:} We mimic the loss of photons by a fictitious beam splitter (BS) in one of the interferometer arms before the phase-shifter [Fig. \ref{EntanglementPhotonLoss}, inset], in order to check whether the phase sensitivity is reduced by the vacuum noise incurred \cite{Oh2017PRA,Huang2023ResPhys}. This loss is given by the BS reflectivity, $\eta_\text{Loss}=\sin^2\theta$; $\tan^2\theta$ being the intensity splitting ratio.  The phase error obtained for thermal input in either SK or CK MZI grows with the loss of photons, but the minimal error still beats the SQL up to 25\% loss in the interferometer arm, thus featuring high resilience to loss. 


\section{Conclusion and discussion}
Imagine a phase microscope illuminated by faint thermal light, say, under water or in space, without a bright coherent light source to serve as a local oscillator or signal. It is impossible at present to achieve sub-shot noise sensitivity with such a device. Yet, we have shown that Kerr-phase elements in interferometry can yield supersensitive phase estimation (SSPE), down to the Heisenberg limit (HL) and even below it, using incoherent, particularly thermal, input. In the absence of loss, the Kerr-nonlinear Mach-Zehnder interferometer (MZI) conserves the total number of photons, and acts on each Fock layer separately. Thus, fluctuations  of the phases between Fock layers do not have any effect on the output intensities, only phases within a Fock layer affect the output photodetection. 
Hence, thermal input states are well suited for SSPE with Kerr-nonlinear MZI, as long as the input phase and amplitude fluctuate slower than the time difference between the interferometer arms. Consequently, the Kerr-nonlinear MZI described here are resilient to low-frequency input noise, allowing the use of unlocked lasers, or other spectrally filtered thermal light sources.

Several important and unexpected results have emerged from our analysis: 
(1) For any degree of nonlinearity and low photon numbers [Eqs. \eqref{PavgThermal},\eqref{DphiThermal}, Fig. \ref{SetupParityDetect}(c)], the $1/\bar{n}$ limit (so called HL) is attained by any input distribution wherein the photon number variance grows as the mean photon number increases, be it pure or mixed.   Remarkably, the achievable phase sensitivity of thermal input surpasses that of pure states in terms of the quantum Fisher information [Eq. \eqref{FQforAll}] for highly efficient photodetection.  For thermal input, SSPE is achievable for detection efficiency down to 0.9 and up to photon loss levels of 25\%.
(2) Even for weak nonlinearity, the Kerr-nonlinear interferometer creates states that can strongly improve phase sensitivity [Fig. \ref{SetupParityDetect}(c)]. (3) While SSPE is achievable for both cross-Kerr (CK) and self-Kerr (SK) nonlinearities, CK provides better sensitivity when the nonlinear strength per photon pair is much smaller than $\pi/2$ [Fig. \ref{SetupParityDetect}(c)].\\ 
 
Comparison with previous SSPE schemes based on Gaussian squeezed states plagued by noise \cite{Monras2006PRA,Pinel2013PRA, Nielsen2023PRL, Aspachs2009phase} shows the advantages of the present Kerr-nonlinear interferometric scheme that is (i) impervious to noise as opposed to squeezed state schemes [Appendix \ref{G}]; (ii) does not require the satisfaction of standard entanglement [Appendix \ref{E}] or nonclassicality criteria [Appendix \ref{F}] and allows information collection by photon-counting or parity measurements without the resources employed by existing schemes \cite{Takeoka2017PRA,Jarzyna2012PRA} [Appendix \ref{H}].

Suitable Kerr-nonlinear MZI may be implemented based on the giant dipole-dipole interaction between counter-propagating few-photon beams in a cloud of cold alkali atoms excited to a Rydberg state \cite{Friedler2005PRA,Shahmoon2011PRA}.  
The scheme invokes electromagnetically induced transparency (EIT) in the slightly-off resonant case.
The detuning can be chosen such that, while both with and without interaction the transmission is the same, the group velocity, and hence, the induced phase, are different depending on the photon numbers in the beam. 
This scheme has been implemented to yield a Kerr nonlinear phase of over $\pi$ for a single photon pair~\cite{Firstenberg2013AttractiveMedium,Drori2023Sc}. Similar two-photon phase shifts have also been demonstrated in the dispersive regime of atom-photon interactions in ultra high-$Q$ cavities \cite{vrajitoarea2020quantum,stolz2022quantum, Bechler2018NatPhys} and tapered fibers \cite{Lodahl2017Nature}.

A Kerr-nonlinear phase element can be used not only as mode coupler within the MZI, but also as a parity filter of the measured output mode [Appendix \ref{B}]. Such filters may bypass the need for photon-number resolved detectors. Another applicaton of Kerr-nonlinear interferometers proposed by us \cite{Opatrny2023ScAdv} has been their use as nonlinear coherent heat machines. 

 The present findings constitute guidelines for nonlinear interferometer implementation that can open new avenues in phase sensing: the ability to achieve SSPE with classical, incoherent, extremely faint light may bring about a paradigm shift in this field.

\begin{acknowledgments}
 The authors acknowledge support from the US-Israel Binational Science Foundation (BSF) and the US National Science Foundation (NSF). EP and OF acknowledge support from the Estate of Louise Yasgour. GK is supported by the Deutsche Forschung Geselschaft (DFG) FOR 2724, by the European commission (FET Open, PATHOS) and by QUANTERA (PACE-IN). TO was
supported by the Czech Science Foundation, grant 20-27994S. NM acknowledges the support of the Feinberg Graduate School (FGS) Dean Postdoctoral Fellowship.
\end{acknowledgments}

\begin{widetext}

\appendix
\renewcommand{\thefigure}{S\arabic{figure}}
\setcounter{figure}{0}

\section{Parity operator mean values in Kerr-nonlinear Mach-Zehnder interferometer (MZI)}\label{A}


\subsection{Self-Kerr (SK) MZI} 
For an input two-mode number state $\ket{n,0}$,  the state right before the final beam splitter [Fig. \ref{SetupParityDetect}(a) of text], is
\begin{align}\label{NumberInputPsiphiAppendix}
\ket{\psi(\phi)}=\frac{i^n}{\sqrt{2}}e^{i\pi/4}(i^{n-1}e^{in\phi}\ket{n,0}+\ket{0,n}).
\end{align}
To find the expectation value of the corresponding parity operator at the output, we apply to the parity operator the beam splitter (BS) transformation
\begin{align}
\Pi_{out}=U_{BS}^\dagger \Pi U_{BS}=\sum_{M} (-i)^M \sum_{k=0}^M (-1)^k\ket{k,M-k}\bra{M-k,k}.
\end{align}
We then have for its expectation value
\begin{align}\label{ParityAvg}
\langle \Pi_{out}\rangle&=\frac{1}{2}((-i)^{n-1} e^{-in\phi}\bra{n,0}+\bra{0,n})\left[\sum_{M}(-i)^M \sum_{k=0}^M (-1)^k\ket{k,M-k}\bra{M-k,k}\right](i^{n-1}e^{in\phi}\ket{n,0}+\ket{0,n})\nonumber\\
&=\frac{1}{2}((-i)^{n-1} e^{-in\phi}\bra{n,0}+\bra{0,n}) \left[(-i)^n i^{n-1} e^{in\phi} \ket{0,n}+(-i)^n (-1)^n  \ket{n,0}\right]\nonumber\\
&=\sin n\phi.
\end{align}

The phase error is then given by
\begin{align}
\Delta\phi=\frac{\sqrt{1-\langle \Pi_{out} \rangle^2}}{\left|\frac{\partial \langle \Pi_{out} \rangle}{\partial \phi}\right|}=\frac{1}{n}.
\end{align}

For coherent state $\ket{\alpha,0}$ input, we get
\begin{align}
\langle \Pi_{out}\rangle
&=e^{-|\alpha|^2}\left[1+\sum_{n=0}^\infty \frac{|\alpha|^{2n}}{n!}\sin n\phi \right]\nonumber\\
&=e^{-|\alpha|^2}\left[1+e^{|\alpha|^2\cos\phi}\sin(|\alpha|^2\sin\phi)\right],
\end{align}
As $\phi\rightarrow 0$, this expression becomes
\begin{align}
\langle \Pi_{out}\rangle\approx e^{-|\alpha|^2}+\sin(|\alpha|^2\phi),
\end{align}
so that for large $|\alpha|^2$
\begin{align}
\Delta\phi\approx\frac{\sqrt{1-e^{-|\alpha|^2}+\sin(|\alpha|^2\phi)}}{\left||\alpha|^2\cos(|\alpha|^2\phi)\right|}\approx \frac{1}{|\alpha|^2}=\frac{1}{\bar{n}}.
\end{align}

For thermal state input, by averaging Eq. \eqref{ParityAvg} over the thermal number-state distribution with average $\bar{n}$, we have  
\begin{align}
\langle \Pi_{out}\rangle= \frac{1}{1+\bar{n}}+\sum_{n=0}^\infty \frac{\bar{n}^n}{(1+\bar{n})^{(n+1)}}\sin n\phi,
\end{align}As $\phi\rightarrow 0$, this expression becomes
\begin{align}
\langle \Pi_{out}\rangle \approx \frac{1}{1+\bar{n}},
\end{align}
and 
\begin{align}
\langle \frac{\partial  \Pi_{out} }{\partial \phi}\rangle \approx \bar{n},
\end{align}
which yields, for large $\bar{n}$, the phase error
\begin{align}
\Delta\phi\approx \frac{\sqrt{1-\frac{1}{(\bar{n}+1)^2}}}{\bar{n}} \approx \frac{1}{\bar{n}}.
\end{align}

For a general input state
\begin{align}
\rho_{in}=\sum_{n}\rho_{nm} \ket{n}\bra{m} \otimes \ket{0} \bra{0},
\end{align}
the output expectation value of the parity operator becomes
\begin{align}
\langle \Pi_{out}\rangle=|\rho_{00}|^2+\sum_{n=0}^\infty |\rho_{nn}|^2 \sin n\phi.
\end{align}

As $\phi\rightarrow 0$, 
\begin{align}
\langle \Pi_{out}\rangle \approx |\rho_{00}|^2,
\end{align}
and 
\begin{align}
\langle \frac{\partial  \Pi_{out} }{\partial \phi}\rangle \approx \bar{n},
\end{align}
which yields for large $\bar{n}$
\begin{align}
\Delta\phi\approx \frac{\sqrt{1-|\rho_{00}|^2}}{ \bar{n} } \approx \frac{1}{ \bar{n}}.
\end{align}

\subsection{Cross-Kerr (CK) MZI}
For input number state $\ket{n,0}$, we have the output state right before the final BS 
\begin{align} 
\ket{\psi(\phi)}\propto \frac{1}{\sqrt{2}}(e^{in\phi}\ket{n,0}+\ket{0,n})~~~ \text{for odd n}.
\end{align}
leading to 
\begin{align}
\langle \Pi \rangle = -i^{n+1} \sin n\phi~~~ \text{for odd n}.
\end{align}
For even $n$, the state $\ket{\psi(\phi)}$ is proportional to a beam splitter transformation, leading to
\begin{align}
\langle \Pi \rangle = \sin^n\phi ~~~~  \text{for even n}.
\end{align}

Therefore, number state with odd $n$ yields the Heisenberg scaling (HL), $\Delta\phi \sim 1/n$, while even $n$ leads to the standard quantum limit (SQL), $\Delta\phi \sim 1/\sqrt{n}$.

For coherent state input, we get
\begin{align}
\langle \Pi_{out} \rangle = e^{-|\alpha|^2}\left[1-\sum_{\text{odd}~ n} \frac{|\alpha|^{2n}}{n!}  i^{n+1} \sin n\phi + \sum_{\text{even}~ n} \frac{|\alpha|^{2n}}{n!} \sin^n\phi \right].
\end{align}

For $\phi\approx\pi/2$ with large $\bar{n}$, we get
\begin{align}
\Delta\phi\approx \frac{1}{|\alpha|^2}.
\end{align}

Similarly, for thermal input, we get
\begin{align}
\langle \Pi_{out} \rangle = \left[\frac{1}{1+\bar{n}}-\sum_{\text{odd}~ n}  \frac{\bar{n}^n}{(1+\bar{n})^{(n+1)}}  i^{n+1} \sin n\phi + \sum_{\text{even}~ n}  \frac{\bar{n}^n}{(1+\bar{n})^{(n+1)}}\sin^n\phi \right],
\end{align}
and the sensitivity for $\phi \approx \pi/2$, for large $\bar{n}$, is then
\begin{align}
\Delta\phi\approx \frac{1}{\bar{n}}.
\end{align}

\begin{figure*}
\begin{center}
\includegraphics[scale=0.42]{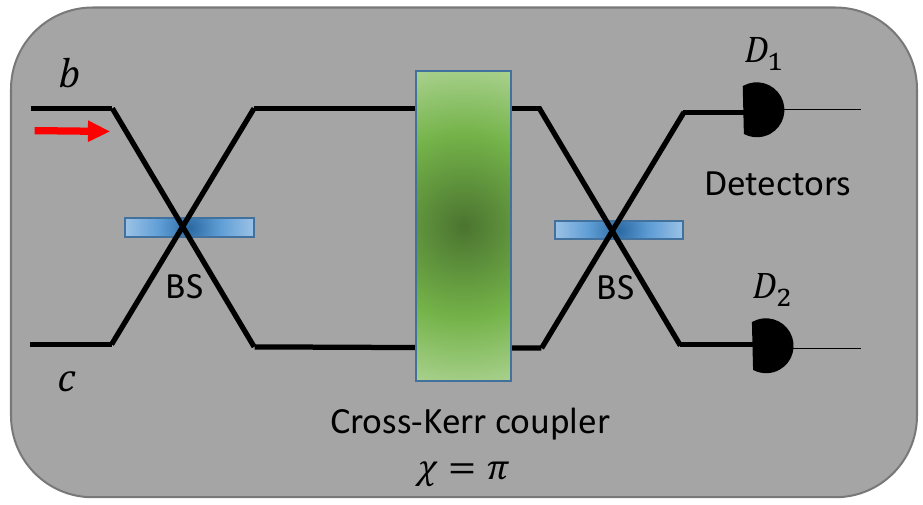}
\end{center}
\caption{ Nonlinear interferometer with cross-Kerr nonlinearity as a parity filter device. }\label{ParityFilter}
\end{figure*}

\section{Cross-Kerr nonlinear MZI as parity measurement device}\label{B}
The parity of the output mode $b_{out}$ [Fig. \ref{SetupParityDetect}(a)] can be measured by sending this mode through a nonlinear interferometer containing cross-Kerr nonlinearity \cite{Gerry2002NonlinearInterferometry}. The nonlinear CK MZI shown in Fig. \ref{ParityFilter}, governed by the unitary operator $\tilde{U}_{CK}=U_{BS} e^{i\chi \hat{n}_b \hat{n}_c} U_{BS}$, transforms the two-mode number state $\ket{n_b,0_c}$, for $\chi=\pi$, as
\begin{align}\label{output}
\tilde{U}_{CK}\ket{n_b,0_c}=i^n\ket{n_1,0_2},~~ \text{for even n}\nonumber\\
\tilde{U}_{CK}\ket{n_b,0_c}=i^n\ket{0_1,n_2}. ~~\text{for odd n}
\end{align}
The subscripts $b$ and $c$ are the input modes, and $1$ and $2$ are the output modes of this device. 
Hence, if the input field contains an even number of photons, then only detector $D_1$ clicks and, only $D_2$ clicks if the input field contains an odd number of photons. The same holds true for an input that is a mixture of $\ket{n_b,0_c}$. 
Thus, the nonlinear interferometer can be used as a parity measurement device merely by checking which detector clicks.

\section{Quantum Fisher Information (QFI) for CK and SK MZI outputs}\label{C}
According to the quantum Cram{\'e}r-Rao bound \cite{cramer1999mathematical}, the optimal phase sensitivity is bounded by
\begin{align}
\Delta\phi_{min}\geq \frac{1}{\sqrt{kF_Q}},
\end{align}
where $k$ is the number of measurements that can be taken to be 1 and  $F_Q$ is the quantum Fisher information that depends neither on the measurement nor on
the estimator and is solely a function of the probing state. 

The input state $\rho_{in}$ is transformed after the phase-shift operation [Fig. \ref{SetupParityDetect}(a) in the text]  to
\begin{align}
\tilde{\rho}(\phi)=U\rho_{in}U^\dagger,
\end{align} 
where $U=U_{PS}(\phi) U_{BS} e^{i\chi a^{\dagger 2} a^2} U_{PS}(\pi/2) U_{BS}$ for SK and $U=U_{PS}(\phi) U_{BS} e^{i\chi a^\dagger a b^\dagger b} U_{PS}(\pi/2) U_{BS}$ for CK.

The corresponding QFI is given by the formula \cite{birrittella2021parity,Toth2014JPA}
\begin{align}\label{FisherInfAppendix}
F_Q(\tilde{\rho})=2\sum_{\substack{k,l \\ \lambda_k + \lambda_l > 0}}\frac{(\lambda_k-\lambda_l)^2}{(\lambda_k+\lambda_l)}|\langle k|a^\dagger a|l\rangle|^2,
\end{align}
where $\{\lambda_k, \ket{k}\}$ are the eigenvalues and their corresponding eigenvectors of $\tilde{\rho}(\phi)$.\\

For a pure state input \cite{Lang2013PRL,Lang2014PRA}, the above expression reduces to 
\begin{align}
F_{Q}=4(\langle \dot \psi(\phi) |\dot \psi(\phi)\rangle-|\langle \dot \psi(\phi) |\psi(\phi) \rangle|^2),
\end{align}
where $|\dot \psi(\phi)\rangle=\frac{\partial}{\partial \phi}\ket{\psi(\phi)}=ia^\dagger a \ket{\psi(\phi)}$. Thus,
\begin{align}
F_{Q}=4(\langle  \psi(\phi) |(a^\dagger a)^2 | \psi(\phi)\rangle-|\langle  \psi(\phi)| a^\dagger a|\psi(\phi) \rangle|^2)=4\langle \Delta a^\dagger a\rangle^2,
\end{align}
whereas for a mixed state, 
\begin{align}\label{FQMixedIneq}
F_Q \leq 4\langle\Delta a^\dagger a\rangle^2.
\end{align}

\subsection{SK interferometer output QFI}
For number-state input $\ket{n,0}$, we get $\ket{\psi(\phi)}$ as given in Eq. \eqref{NumberInputPsiphiAppendix}, and the expectation values are
\begin{align}
\langle  \psi(\phi) |(a^\dagger a)^2 | \psi(\phi)\rangle =  \frac{1}{2}n^2,\langle  \psi(\phi) |(a^\dagger a) | \psi(\phi)\rangle = \frac{1}{2}n.
\end{align}
Using these expectation values, we get
\begin{align}\label{FQnumber}
(F_Q)_{number}=n^2,~~~~ \Delta\phi_{min}=\frac{1}{\sqrt{F_Q}}=\frac{1}{n}.
\end{align}

Similarly, for coherent-state input
\begin{align}
\langle  \psi(\phi) |(a^\dagger a)^2 | \psi(\phi)\rangle = \frac{1}{2}|\alpha|^4+\frac{1}{2}|\alpha|^2,
\langle  \psi(\phi) |(a^\dagger a) | \psi(\phi)\rangle = \frac{1}{2}|\alpha|^2,
\end{align}
and 
\begin{align}\label{FQcoherent}
(F_Q)_{coherent}=|\alpha|^4+2|\alpha|^2=\bar{n}^2+2\bar{n},~~~~ \Delta\phi_{min}=\frac{1}{\sqrt{F_Q}}=\frac{1}{\sqrt{\bar{n}^2+2\bar{n}}}.
\end{align}

For thermal-state input, using Eq. \eqref{FisherInfAppendix}, we get
\begin{align}\label{FQthermal}
(F_Q)_{thermal}=2\bar{n}^2+\bar{n},~~~~ \Delta\phi_{min}=\frac{1}{\sqrt{F_Q}}=\frac{1}{\sqrt{2\bar{n}^2+\bar{n}}}.
\end{align}

Our results do not depend on the choice of the phase generator Hamiltonian \cite{Jarzyna2012PRA}. This can be shown by comparing the QFI obtained for the two-mode phase-generator Hamiltonian $H_{2mode}=\phi N_d/2$, where $N_d=a^\dagger a-b^\dagger b$ is the two-mode number difference operator, with that obtained for the single-mode (single arm) phase generator Hamiltonian, $H_{1mode}=\phi a^\dagger a$ in our Kerr-nonlinear MZI scheme.

It is straightforward to show that $H_{2mode}$ yields the same form of $F_Q$ for pure- or mixed-state input as Eqs. (\ref{FisherInfAppendix}-\ref{FQMixedIneq}) upon replacing $a^\dagger a$ by $N_d/2$.

In the self-Kerr (SK) interferometer with $\chi=\pi/2$ and $H_{2mode}$, we then obtain the following:\\
\begin{subequations}
For number-state input,
\begin{align}\label{FpQnumber}
\langle N_d^2 \rangle=n^2, \langle N_d \rangle=0; (F'_Q)_{number}=n^2;
\end{align}
for coherent-state input,
\begin{align}\label{FpQcoherent}
\langle N_d^2 \rangle=\bar{n}^2+\bar{n}, \langle N_d \rangle=0; (F'_Q)_{coherent}=\bar{n}^2+\bar{n};
\end{align}
and for thermal-state input, 
\begin{align}\label{FpQthermal}
(F'_Q)_{thermal}=2\bar{n}^2+\bar{n}.
\end{align}
\end{subequations}   
Comparing these results with Eq. (\ref{FQforAll}a) of the main text, we find that the above QFI values are the same as those obtained with the single-mode phase generator $H_{1mode}$ for number-state and thermal-state inputs. The QFI for the coherent-state input is different for the two phase generators, but for both the phase sensitivity still surpasses the HL. 
Importantly, thermal state input exhibits QFI higher than that of the coherent state input, be it an interferometer with single-mode or two-mode phase generator. Similar conclusions hold for the cross-Kerr (CK) interferometer.  

Thus, the QFI in SK/CK MZI for thermal- and number-state input is independent of the choice of the phase generators. Therefore, a local oscillator is not needed for phase reference in the CK/SK MZI scheme.

\begin{figure*}[h]
\begin{center}
\includegraphics[scale=0.4]{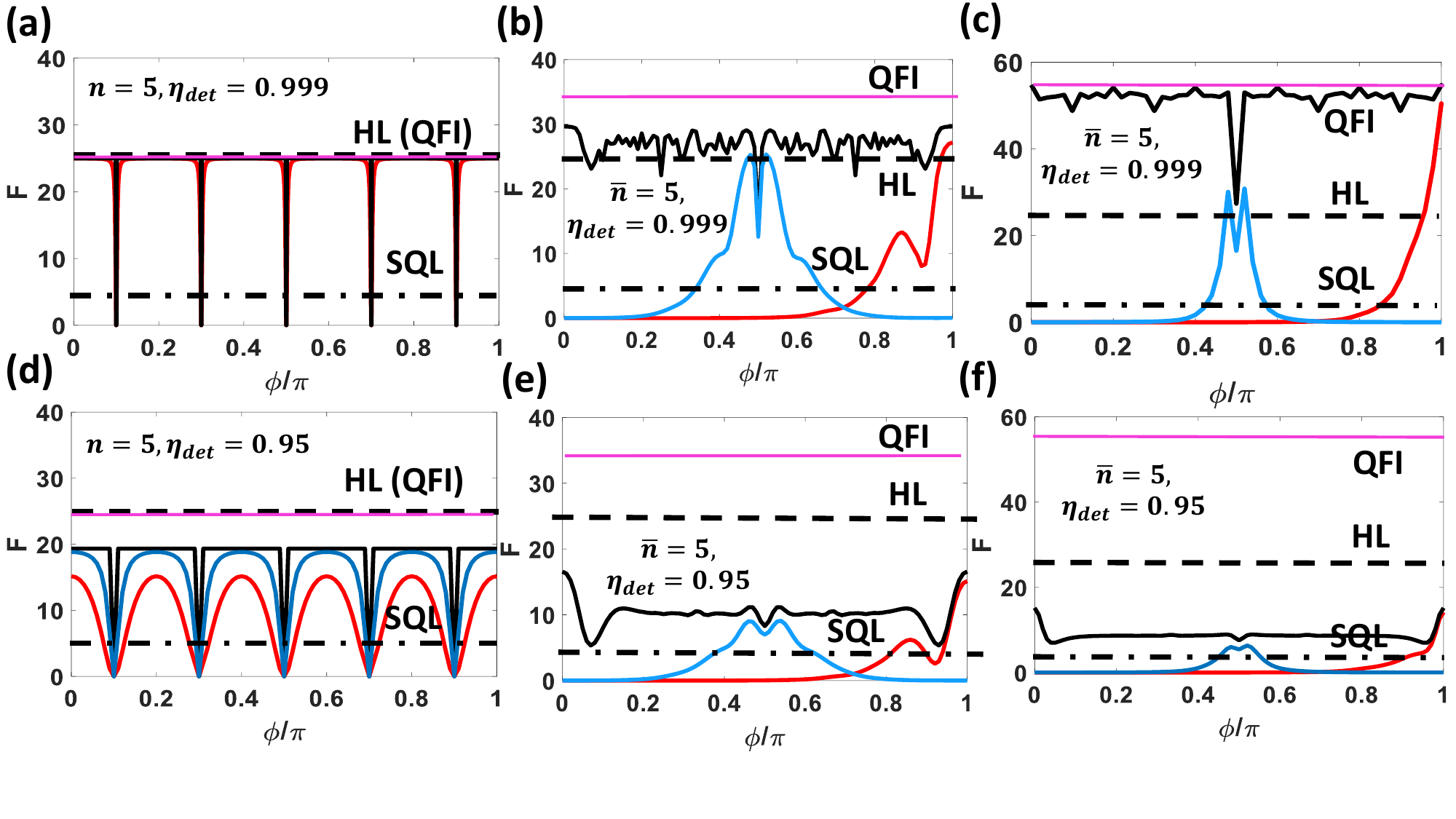}
\end{center}
\caption{ Fisher information for SK MZI ($\chi=\pi/2$) obtained by photon-number resolving detectors: (a) number-state ($n=5$) input with detector efficiency $\eta_\text{det}= 0.999$, (b) coherent state ($\bar{n}=5)$ input with $\eta_\text{det}= 0.999$,  (c) thermal state ($\bar{n}=5)$ input with $\eta_\text{det}= 0.999$, (d) number-state ($n=5$) input with $\eta_\text{det}=0.95$, (e) coherent state ($\bar{n}=5)$ input with $\eta_\text{det}= 0.95$ and (f) thermal state ($\bar{n}=5)$ input with $\eta_\text{det}= 0.95$. The SQL (dot-dashed line) here corresponds to $F_{SQL}=\bar{n}=5$ and the HL (dashed line) to $F_{HL}=\bar{n}^2=25$. The QFI (magenta line) for thermal input is 55, for coherent input is 35 and for number input is 25. Red thick line: Single-detector photodetection, Blue thick line: intensity-difference detection, Black thick line: photodetection by both detectors.       }\label{CoherentInefficient}
\end{figure*}

\subsection{CK interferometer output QFI}
For number-state input  $\ket{n,0}$, 
\begin{align}
&\langle  \psi(\phi) |(a^\dagger a)^2 | \psi(\phi)\rangle =  \frac{1}{4}(n^2+n)~~ \text{for even N},\\
&\langle  \psi(\phi) |(a^\dagger a)^2 | \psi(\phi)\rangle =  \frac{1}{2}n^2 ~~\text{for odd N},\\
&\langle  \psi(\phi) |(a^\dagger a) | \psi(\phi)\rangle = \frac{1}{2}n,
\end{align}
and 
\begin{align}
&(F_Q)_{number}=n^2,~~~~ \Delta\phi_{min}=\frac{1}{\sqrt{F_Q}}=\frac{1}{n}, ~~\text{for odd N}\\ &(F_Q)_{number}=n,~~~~\Delta\phi_{min}=\frac{1}{\sqrt{F_Q}}=\frac{1}{\sqrt{n}}, ~~~ \text{for even N}.
\end{align}

For coherent-state input,
\begin{align}
\langle  \psi(\phi) |(a^\dagger a)^2 | \psi(\phi)\rangle = \frac{3}{8}|\alpha|^4+\frac{1}{2}|\alpha|^2,
\langle  \psi(\phi) |(a^\dagger a) | \psi(\phi)\rangle = \frac{1}{2}|\alpha|^2,
\end{align}
and 
\begin{align}
(F_Q)_{coherent}=\frac{1}{2}|\alpha|^4+2|\alpha|^2=\frac{1}{2}\bar{n}^2+2\bar{n},~~~~ \Delta\phi_{min}=\frac{1}{\sqrt{F_Q}}=\frac{1}{\sqrt{\frac{1}{2}\bar{n}^2+2\bar{n}}}.
\end{align}

For thermal-state input (from Eq. \eqref{FisherInfAppendix}), we get
\begin{align}
(F_Q)_{thermal}=\bar{n}^2+\bar{n},~~~~ \Delta\phi_{min}=\frac{1}{\sqrt{F_Q}}=\frac{1}{\sqrt{\bar{n}^2+\bar{n}}}.
\end{align}

\begin{figure*}[h]
\begin{center}
\includegraphics[scale=0.4]{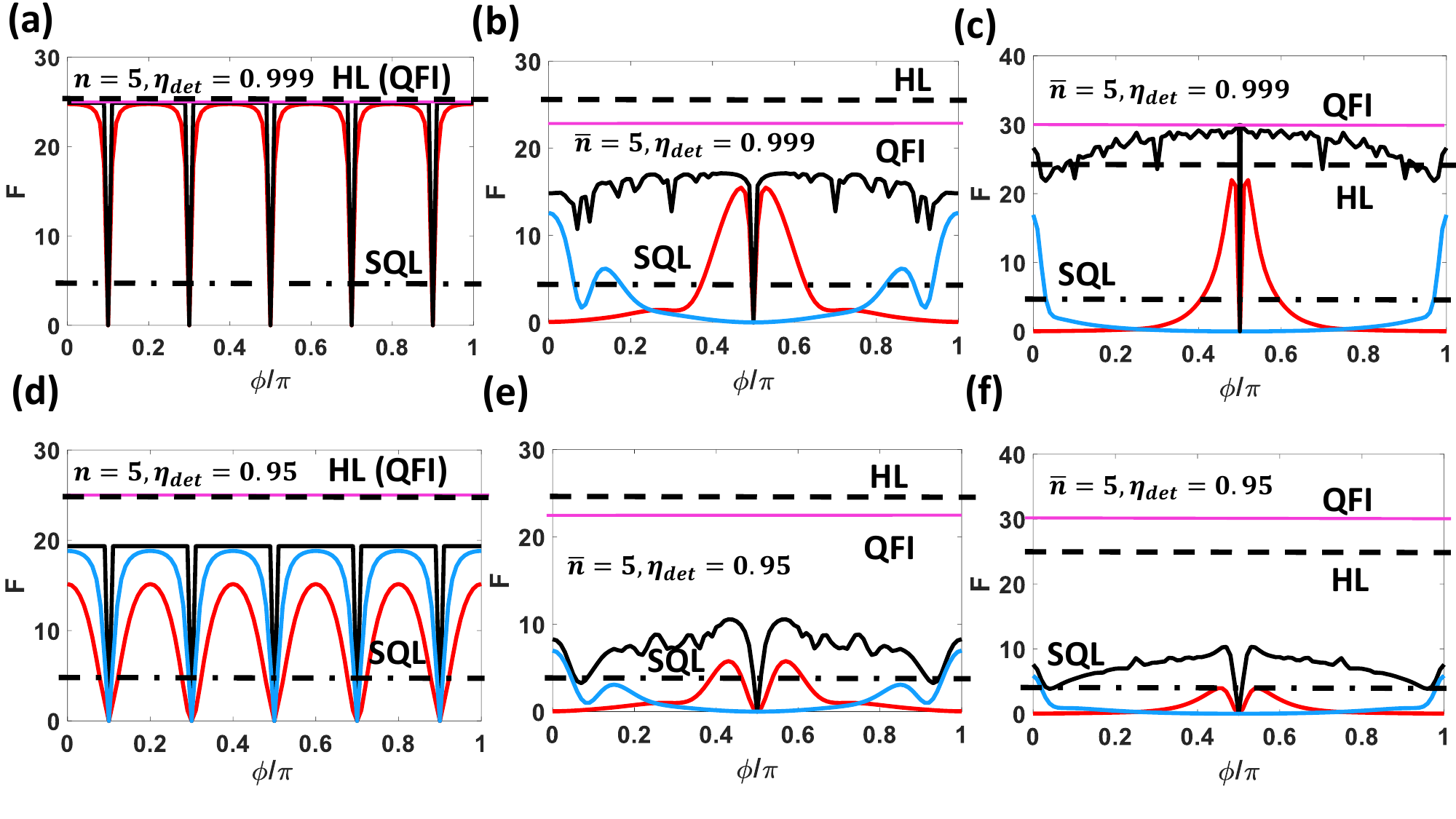}
\end{center}
\caption{ Fisher information for CK MZI ($\chi=\pi/2$) obtained by photon-number resolving detectors: (a) number-state ($n=5$) input with detector efficiency $\eta_\text{det}= 0.999$, (b) coherent state ($\bar{n}=5)$ input with $\eta_\text{det}= 0.999$,  (c) thermal state ($\bar{n}=5)$ input with $\eta_\text{det}= 0.999$, (d) number-state ($n=5$) input with $\eta_\text{det}=0.95$, (e) coherent state ($\bar{n}=5)$ input with $\eta_\text{det}= 0.95$ and (f) thermal state ($\bar{n}=5)$ input with $\eta_\text{det}= 0.95$. The SQL (dot-dashed line) here corresponds to $F_{SQL}=\bar{n}=5$ and the HL (dashed line) to $F_{HL}=\bar{n}^2=25$. The QFI (magenta line) for thermal input is 30, for coherent input is 22.5 and for number input is 25. Red thick line: Single-detector photodetection, Blue thick line: intensity-difference detection, Black thick line: photodetection by both detectors.       }\label{CoherentInefficientCK}
\end{figure*}

\begin{figure*}[h!]
\begin{center}
\includegraphics[scale=0.25]{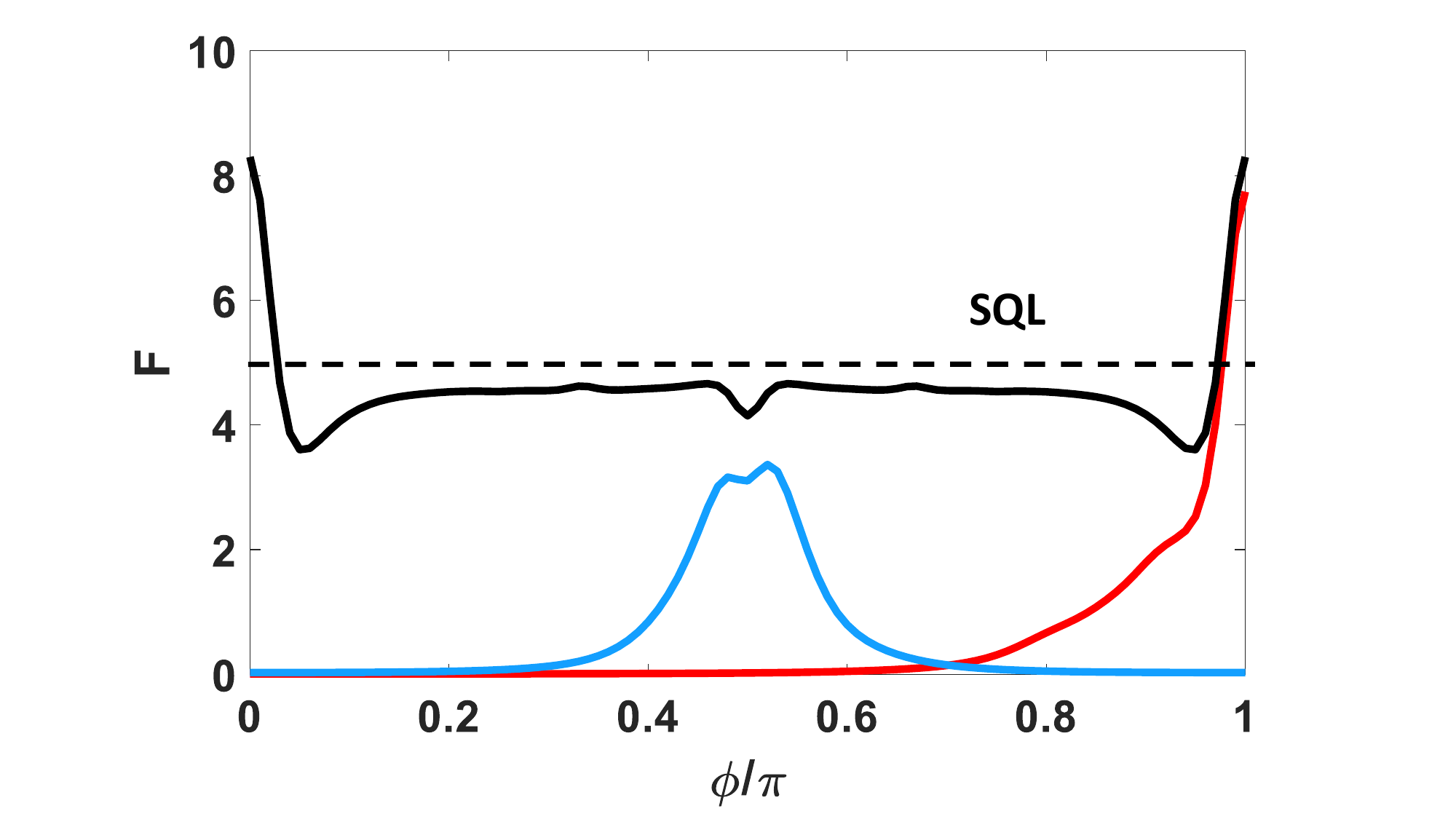}
\end{center}
\caption{ Fisher information for SK MZI with thermal state ($\bar{n}=5)$ input for $\chi=\pi/2$ and  detector efficiency $\eta_\text{det}= 0.9$.  The SQL (dashed line) here corresponds to $F_{SQL}=\bar{n}=5$. Red thick line: Single-detector photodetection, Blue thick line: intensity-difference detection, Black thick line: photodetection on both detectors.       }\label{ThermalDetEff09}
\end{figure*}

\begin{figure*}[h!]
\begin{center}
\includegraphics[scale=0.37]{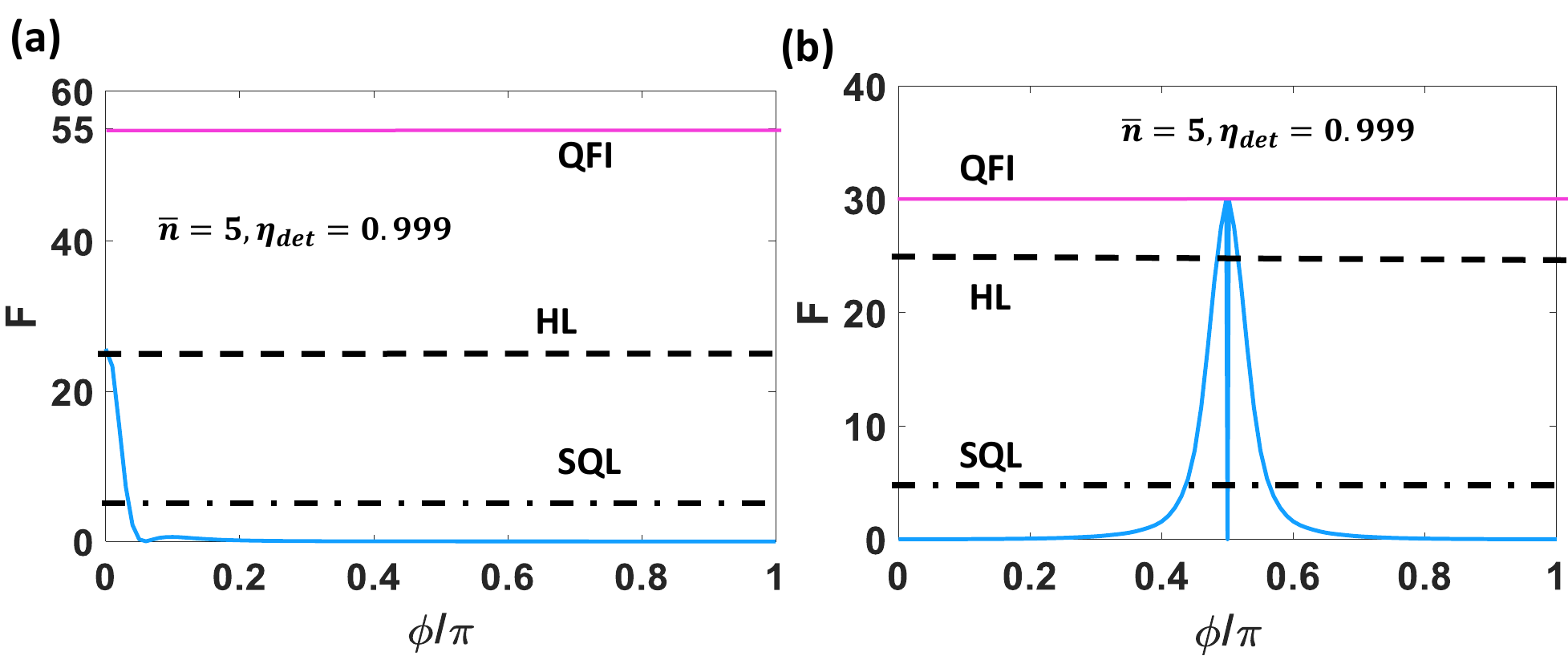}
\end{center}
\caption{ (a) Fisher information, based on parity measurement, for SK MZI with thermal state ($\bar{n}=5)$ input for $\chi=\pi/2$ and  detector efficiency $\eta_\text{det}= 0.999$. (b) Similarly, for CK MZI. The SQL (dot-dashed line) here corresponds to $F_{SQL}=\bar{n}=5$ and the HL (dashed line) to $F_{HL}=\bar{n}^2=25$. The QFI (magenta line) for thermal input in SK MZI is 55 and in CK MZI is 30.         }\label{ParityThermal}
\end{figure*}

\section{Photodetection efficiency effects}\label{D}
With perfect detectors $\eta_\text{det}=1$, photodetection on both output channels saturates the QFI limit for thermal and number inputs while not for coherent input [Figs. \ref{CoherentInefficient} and \ref{CoherentInefficientCK}]. With near-perfect detectors, mean photon numbers being equal, the highest $F$ is reached by thermal states, which then beat the HL and  surpass coherent and Fock states. For  $\eta_\text{det}\leq 0.97$ the order is reversed: the highest $F$ is reached by Fock states and the lowest by thermal states. Yet, thermal input still beats the SQL for $\eta_{\text{det}} \geq 0.9$ [Fig. \ref{ThermalDetEff09}]. We note that the FI, based on parity measurement, achieves the QFI for CK MZI  and not for SK MZI [Fig. \ref{ParityThermal}].

\section{Sufficient conditions for two-mode entanglement in Kerr-nonlinear MZI}\label{E}
To check whether the two-mode output states (after the second beam splitter of the setup given in Fig. \ref{SetupParityDetect}(a) of the text) satisfy the standard sufficient conditions of entanglement, we consider the following quantities
\begin{align}
E_{HZ}&=\langle \hat n_a \hat n_b \rangle_{out}-|\langle a_{out}^\dagger b_{out} \rangle|^2,\\
E_{SV}&=\left|\begin{array}{ccc}
1 & \langle a_{out}^\dagger \rangle & \langle b_{out}^\dagger \rangle\\
\langle a_{out} \rangle & \langle a_{out}^\dagger a_{out} \rangle & \langle a_{out}^\dagger b_{out}^\dagger \rangle\\
\langle b_{out} \rangle & \langle a_{out} b_{out} \rangle & \langle b_{out}^\dagger b_{out} \rangle
\end{array}\right|.
\end{align} 
These quantities are constructed based on the Hillery-Zubairy \cite{Hillery2006PRL}, and Shchukin-Vogels entanglement criteria \cite{Shchukin2005PRL}, respectively. The negativity of these quantities for a two-mode state is a sufficient condition for entanglement. As shown in Fig. \ref{Entanglement}(a), this condition is not satisfied by output states in Kerr-nonlinear interferometers when the input is a thermal state. 

\begin{figure}
\begin{center}
\includegraphics[scale=0.4]{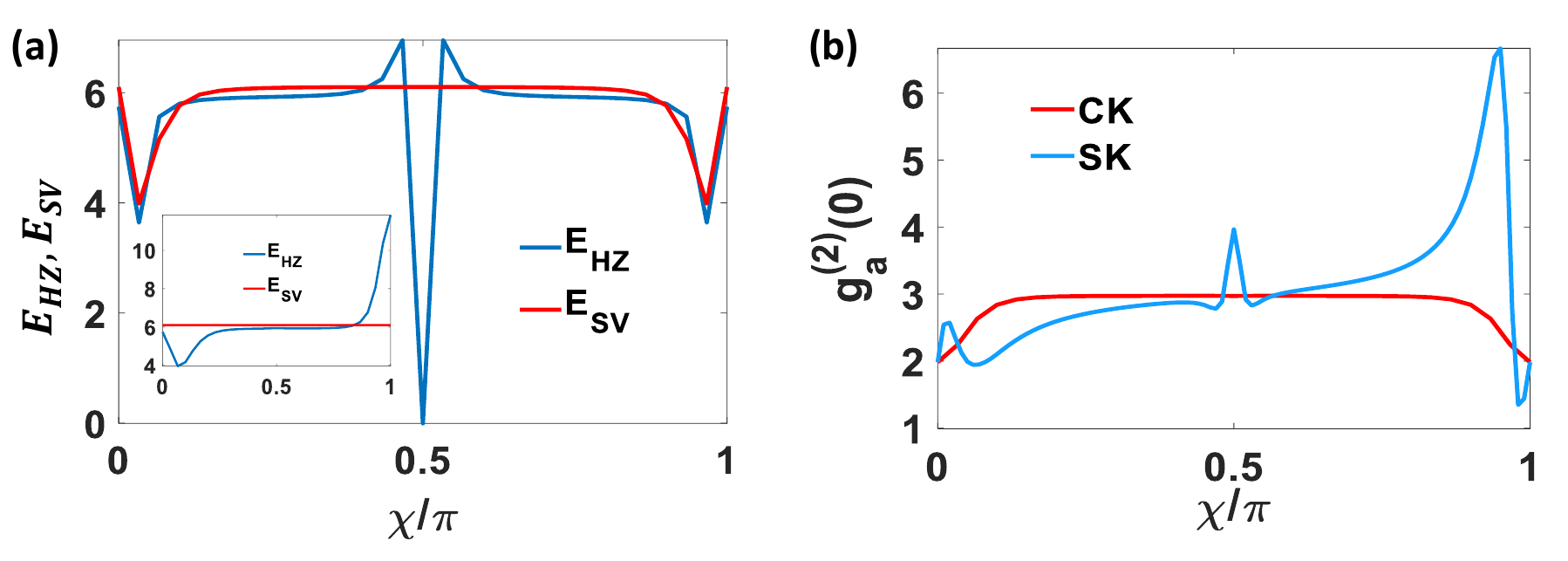}
\end{center}
\caption{(a) Sufficient entanglement conditions, characterized by the negativity of $E_{HZ}$ and $E_{SV}$ for the two-mode states after the second beam splitter of the SK MZI, are \textit{not satisfied} for thermal input for any nonlinear Kerr strength $\chi$. Here $\bar{n}_a$=5. Inset: Similarly for CK MZI. (b) The zero time-delay second-order coherence function $g^{(2)}_{a}(0)$ of the mode $a$ after the second beam splitter as a function of $\chi$ for thermal input ($\bar{n}_a=5$) does not feature sub-Poissonian [$g^{(2)}_{a}(0)<1]$ photon statistics. }
\label{Entanglement}
\end{figure}

\section{Sufficient condition for single-mode nonclassicality}\label{F}
A single-mode field state is nonclassical if the zero time-delay second-order coherence function, defined by 
\begin{align}
g^{(2)}(0)=\frac{\langle a^{\dagger 2}a^2 \rangle}{\langle a^\dagger a\rangle},
\end{align}
becomes less than 1 (sub-Poissonian). 

We find that none of the output modes exhibit sub-Poissonian ($g^{(2)}_{a}(0)<1)$ photon statistics, which is a nonclassical characteristic of the field, as seen in Fig. \ref{Entanglement}(b).

\section{Comparison of state preparation schemes and their QFI}\label{G}
Quantum states with undefined number of photons are known to allow phase estimation  with a sensitivity that surpasses the Heisenberg limit (HL).  The few quantum states known to allow such supersensitivity  are  all Gaussian, and among them, squeezed vacuum states are the optimal \cite{Monras2006PRA}. However, experimental generation of squeezed vacuum states with large photon numbers, without any noise, is an inherently insurmountable challenge \cite{Vahlbruch2008PRL, Nielsen2023PRL} because the generated state has inevitably extra noise in its anti-squeezing quadrature \cite{Yu2020PRAppl}.

By contrast, in our case, the probe state generated from thermal input in an interferometer with either cross-Kerr (CK) or self-Kerr (SK) nonlinearity is very noisy but this does not preclude its ability to surpass the HL.

Experimentally, the challenge has been for years to achieve large Kerr nonlinearity per photon pair, but this challenge has been overcome by our team in a Rydberg polariton setup \cite{Firstenberg2013AttractiveMedium, peyronel2012quantum,Drori2023Sc} and by others in circuit QED setups \cite{vrajitoarea2020quantum}. A large mean photon number does not present an experimental challenge for Kerr nonlinear MZI, and our results show that, if needed, bright noisy sources with  $\bar{n}\gg 1$ can be employed with the same scaling of the quantum Fisher information (QFI) as per Eqs. (\ref{FQforAll}a) and (\ref{FQforAll}b) in the text.

\begin{figure}
\begin{center}
\includegraphics[scale={0.4}]{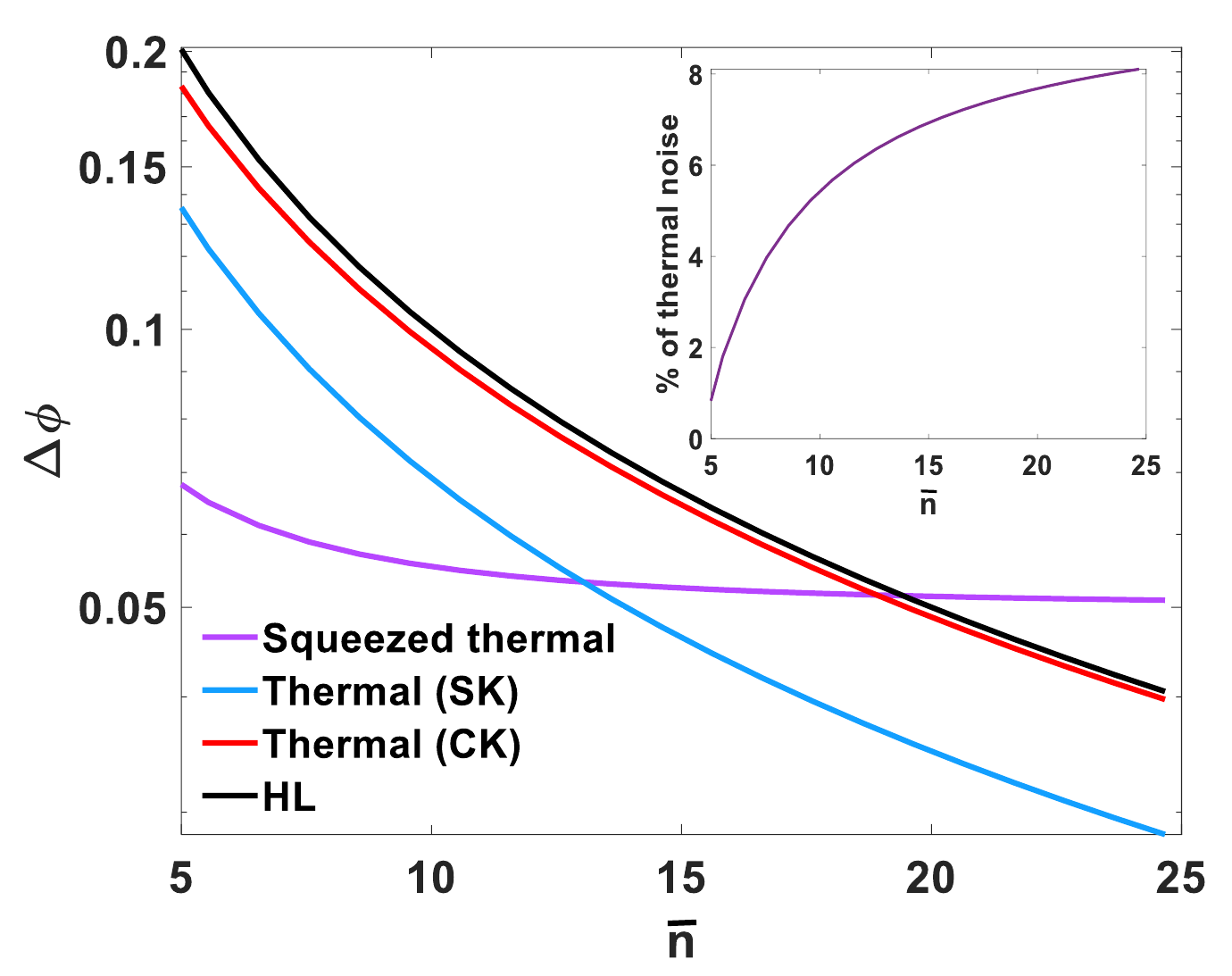}
\end{center}
\caption{The phase sensitivity $\Delta\phi=1/\sqrt{F_Q}$ of CK/SK MZI fed by thermal noise compared with that of the squeezed thermal obtained from Eq. \eqref{SqThermal}. The inset in the figure shows the percentage of thermal noise in the squeezed thermal state. Namely, in Eq. \eqref{SqThermalmean} we fix $n_r$ (the squeezed part), say, at $n_r=4.5$, and increase $n_b$ (the thermal noise), thus increasing $\bar{n}$. When the percentage of thermal noise, $n_b/\bar{n}\times 100$, exceeds 6\% (7\%), the QFI values of the SK (CK) interferometers are larger than that obtained in Eq. \eqref{SqThermal} with the squeezed thermal state having the same mean photon number. Since our scheme can be fed by purely thermal noise, it is therefore advantageous when thermal noise is appreciable.  }
\label{FSqThSKCK}
\end{figure}

We compare QFI achievable by thermal noise in our scheme with that of squeezed states mixed with thermal noise, which realistically occurs in experiments \cite{Yu2020PRAppl}.  For a  squeezed thermal  state \cite{Aspachs2009phase}, the QFI is 
\begin{align}\label{SqThermal}
(F_Q)_{SqThermal}=16 n_r (n_r+1)\left(1+\frac{1}{(2n_b+1)^2}\right)^{-1},
\end{align}
where $n_r$ is the mean photon number associated with squeezing and $n_b$ is the corresponding contribution of thermal noise, the total mean photon number being
\begin{align}\label{SqThermalmean}
\bar{n}=n_b+(2n_b+1) n_r.
\end{align}
In Fig. \ref{FSqThSKCK}, $\Delta\phi=1/\sqrt{F_Q}$ is plotted as a function of $\bar{n}$ and compared with our results whereby
\begin{align}
(F_Q)_{thermal}&=(2 \bar{n}^2+\bar{n}), ~~~     \text{for SK}\\
(F_Q )_{thermal}&=(\bar{n}^2+\bar{n}). ~~~     \text{for CK}
\end{align}
This figure shows that when the thermal noise exceeds more than 6 (7) percent of the total intensity, the phase sensitivity achieved by the state generated in an SK (CK) interferometer from thermal noise input outperforms that of the squeezed thermal state.

Thus, state preparation by our Kerr-nonlinear scheme has a salient metrological advantage compared to Gaussian squeezed states.

\section{Comparison of measurement schemes}\label{H}
The best possible measurement schemes currently  employed to surpass the  HL using squeezed vacuum states are either positive-operator valued measurement (POVM) or homodyne measurements \cite{Monras2006PRA, Nielsen2023PRL} 

In the case of POVM, finding the optimal positive operators for attaining such a precision is a highly nontrivial undertaking. Even if they happen to be known, the optimal POVM depends on the true value of the phase to be estimated \cite{Monras2006PRA}. Therefore, the measurement at a given step relies on the data from  previous steps.

In a homodyne measurement scheme, measuring a quadrature operator $\hat{X}$   for squeezed vacuum input does not yield any information about the phase, because $\langle \hat{X}\rangle=0$. One may choose  $\hat{X}^2$  as an estimator. However, the optimal phase to be estimated then becomes dependent on the input photon number \cite{Nielsen2023PRL}. As a result, for a given input photon number, one needs to set apriori the phase to be estimated. Alternatively,  Bayesian estimation process requires to estimate the phase from the data obtained by measuring  many samples \cite{Nielsen2023PRL}.  Yet, due to the intrinsic nature of squeezed vacuum, homodyne measurements can attain the HL and surpass it only for a small range of phases.

By contrast, the measurement techniques that we invoke to attain the quantum Cram{\'e}r-Rao bound are photon-count and parity measurements. The Fisher information for these measurements becomes almost independent of the phase to be estimated for all photon numbers.

\end{widetext}



%

\end{document}